\documentclass{nature}

\usepackage[%showframe,%Uncomment any one of the following lines to test 
margin=0.67in,
]{geometry}

\usepackage{graphicx}
\usepackage{multicol}
\makeatletter
\let\saved@includegraphics\includegraphics
\AtBeginDocument{\let\includegraphics\saved@includegraphics}
\renewenvironment*{figure}{\@float{figure}}{\end@float}
\makeatother
\usepackage{subcaption} 

\usepackage[utf8]{inputenc}
\usepackage[english]{babel}
\usepackage[T1]{fontenc}
\usepackage{amsmath}
\makeatletter
\newcommand{\pushright}[1]{\ifmeasuring@#1\else\omit\hfill$\displaystyle#1$\fi\ignorespaces}
\newcommand{\pushleft}[1]{\ifmeasuring@#1\else\omit$\displaystyle#1$\hfill\fi\ignorespaces}
\makeatother

\usepackage{tikz}
\usepackage{amssymb}
\usepackage{color}
\usepackage{mathrsfs}
\usepackage{amsbsy}
\usepackage{amsthm}
\usepackage{float}
\usepackage{balance}
\usepackage{thmtools,thm-restate}
\usepackage{graphicx}
\usepackage{tikz}

\usepackage{comment}

\usepackage{bbm}
\usepackage{bm}
\usepackage{epsfig}
\usepackage{xfrac}
\usepackage{xcolor}
\usepackage{enumerate}
\usepackage[shortlabels]{enumitem}
\usepackage{graphicx}% Include figure files
\usepackage{dcolumn}% Align table columns on decimal local
\usepackage{bm}% bold math
\usepackage{hyperref}
\hypersetup{
colorlinks=true,
linkcolor=blue,
filecolor=blue,
citecolor=blue,  
urlcolor=blue,
}

\newtheorem{theorem}{Theorem}

\newtheorem{corollary}[theorem]{Corollary}
\usepackage{braket}
\renewcommand{\gtrsim}{\raisebox{-0.13cm}{~\shortstack{$>$ \\[-0.09cm]$\sim$}}~}

\renewcommand{\v}[1]{\ensuremath{\mathbf{#1}}} % for vectors
\newcommand{\abs}[1]{\left| #1 \right|} % for absolute value
\newcommand{\norm}[1]{\left\| #1 \right\|} % for norm
\newcommand{\trace}{\mathrm{Tr}}
\newcommand{\poly}{\mathrm{poly}}

\newcommand{\group}{{\mathrm{global}}} % for Laplacian
\newcommand{\point}{{\mathrm{local}}}
\newcommand{\charge}{{\mathrm{charge}}}

\newcommand{\txr}{{\mathscr{R}}}

\newcommand{\mC}{{\mathcal{C}}}

\newcommand{\mU}{{\mathcal{U}}}

\newcommand{\mN}{{\mathcal{N}}}

\newcommand{\mR}{{\mathcal{R}}}
\newcommand{\frakF}{{\mathfrak{F}}}
\newcommand{\frakc}{{\mathfrak{c}}}

\newcommand{\mE}{{\mathcal{E}}}

\newcommand{\id}{{\mathbbm{1}}}

\newcommand{\barF}{{F^\infty}}
\newcommand{\barP}{\overline{P}}

\newcommand{\bepsilon}{\overline{\epsilon}}
\newcommand{\bdelta}{\overline{\delta}}

\newcommand{\rep}{{\mathrm{rep}}}

\newcommand{\frakJ}{{\mathfrak{J}}}

\newcommand{\bR}{{\mathbb{R}}}
\newcommand{\bZ}{{\mathbb{Z}}}

\newcommand{\cov}{{\mathrm{cov}}}
\newcommand{\supp}{{\mathrm{supp}}}
\renewcommand{\epsilon}{\varepsilon}
\newcommand{\appropto}{\mathrel{\vcenter{
  \offinterlineskip\halign{\hfil$##$\cr
    \propto\cr\noalign{\kern2pt}\sim\cr\noalign{\kern-2pt}}}}}

\newcommand{\LtoS}{{S\leftarrow L}}
\newcommand{\StoL}{{L\leftarrow S}}
\newcommand{\AtoB}{{B\leftarrow A}}

\newcommand{\CtoSA}{{SA\leftarrow C}}
\newcommand{\SAtoC}{{C \leftarrow SA}}
\newcommand{\CtoLA}{{LA \leftarrow C}}
\newcommand{\LAtoC}{{C \leftarrow LA}}

\let\baraccent=\= % rename builtin command \= to \baraccent
\renewcommand{\=}[1]{\stackrel{#1}{=}} % for putting numbers above =
\newcommand{\thmref}[1]{\hyperref[#1]{Theorem~\ref{#1}}}
\newcommand{\lemmaref}[1]{\hyperref[#1]{Lemma~\ref{#1}}}
\newcommand{\corollaryref}[1]{\hyperref[#1]{Corollary~\ref{#1}}}
\newcommand{\figref}[1]{\hyperref[#1]{Fig.~\ref{#1}}}
\newcommand{\tableref}[1]{\hyperref[#1]{Table~\ref{#1}}}
\newcommand{\figaref}[1]{\hyperref[#1]{Fig.~\ref{#1}a}}
\newcommand{\figbref}[1]{\hyperref[#1]{Fig.~\ref{#1}b}}
\newcommand{\figcref}[1]{\hyperref[#1]{Fig.~\ref{#1}c}}
\newcommand{\figdref}[1]{\hyperref[#1]{Fig.~\ref{#1}d}}
\newcommand{\figeref}[1]{\hyperref[#1]{Fig.~\ref{#1}e}}
\renewcommand{\eqref}[1]{\hyperref[#1]{Eq.~(\ref{#1})}}
\newcommand{\secref}[1]{\hyperref[#1]{Sec.~\ref{#1}}}
\newcommand{\eqsref}[2]{\hyperref[#1]{Eqs.~(\ref{#1})-(\ref{#2})}}
\newcommand{\appref}[1]{\hyperref[#1]{Appx.~\ref{#1}}}

\title{Approximate symmetries and quantum error correction}

\author{Zi-Wen Liu$^{1,2*}$, Sisi Zhou$^{1,3,4*}$}

\begin{document}

\maketitle

\begin{affiliations}
 \item Perimeter Institute for Theoretical Physics, Waterloo, Ontario N2L 2Y5, Canada
 \item Yau Mathematical Sciences Center, Tsinghua University, Beijing 100084, China
 \item Institute for Quantum Information and Matter, California Institute of Technology, Pasadena, CA 91125, USA
 \item Pritzker School of Molecular Engineering, The University of Chicago, Illinois 60637, USA
 \vspace{0.1in}
 \item[*] Correspondence and requests for materials
should be addressed to Z.-W.L. (email: zwliu0@mail.tsinghua.edu.cn) and S.Z. (email: sisi.zhou26@gmail.com). The author names are in alphabetical order. 
\end{affiliations}

\begin{abstract}

Quantum error correction (QEC) is a key concept in quantum computation as well as many areas of physics. There are fundamental tensions between continuous symmetries and QEC. One vital situation is unfolded by the Eastin--Knill theorem, which forbids the existence of QEC codes that admit transversal continuous symmetry actions (transformations). Here, we systematically study the competition between continuous symmetries and QEC in a quantitative manner. We first define a series of meaningful measures of approximate symmetries motivated from different perspectives, and then establish a series of trade-off bounds between them and QEC accuracy utilizing multiple different methods. Remarkably, the results allow us to derive general quantitative limitations of transversally implementable logical gates, an important topic in fault-tolerant quantum computation. As concrete examples, we showcase two explicit types of quantum codes, obtained from quantum Reed--Muller codes and thermodynamic codes, respectively, that nearly saturate our bounds. Finally, we discuss several potential applications of our results in physics. 
\end{abstract}

\vspace{0.1in}


\newpage

%\begin{multicols}{1}

\section*{Introduction}

Symmetry has long been a pivotal concept and tool in physics.  Continuous symmetries are those described by transformations that vary continuously as a function of some parameterization, mathematically modeled by Lie groups.   Associated with conservation laws as dictated by the celebrated Noether's theorem\cite{noether2005invariant}, 
  continuous symmetries 
 ubiquitously arise in physical systems and play fundamental roles in their behaviors.  A basic, prototypical example of a continuous symmetry group is $U(1)$, coming with a conserved quantity that may correspond to charge, particle number, energy, etc., depending on the physical scenario. 
 
\begin{figure}[H]
\centering
\includegraphics[width=0.7\textwidth]{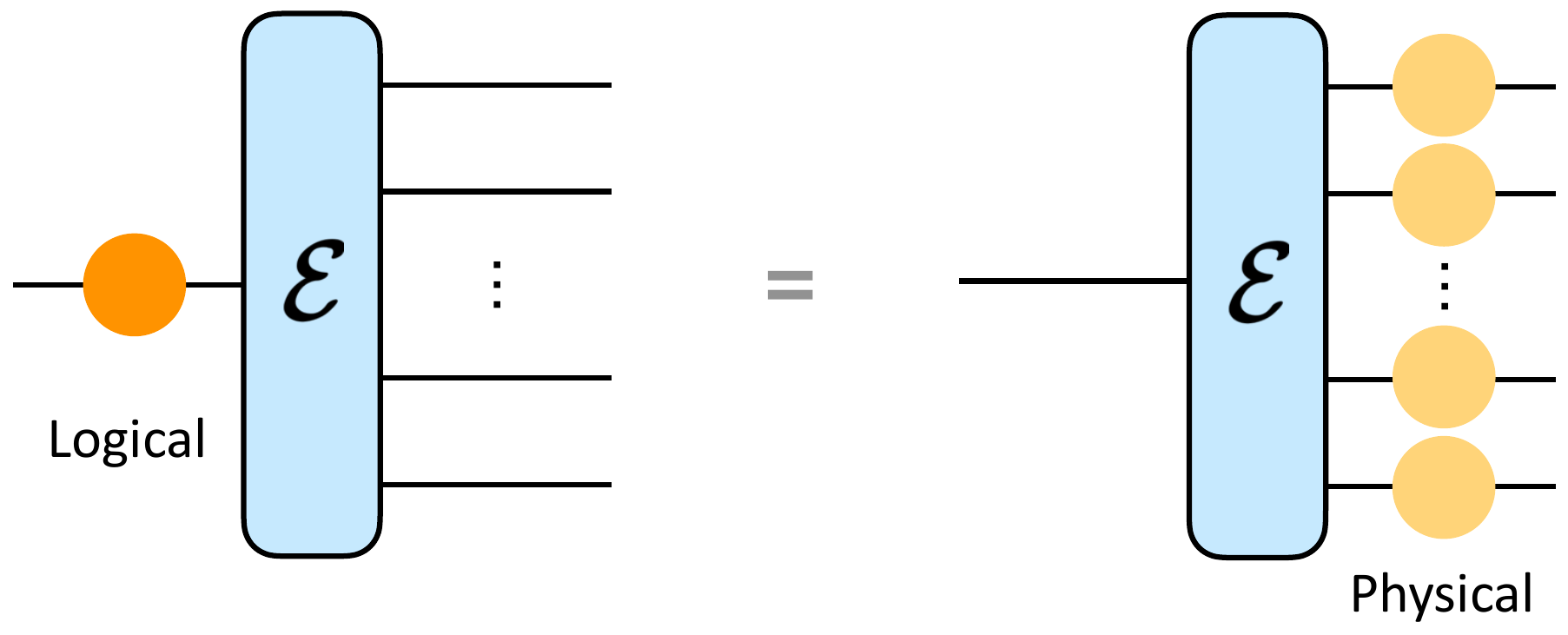}
 \caption{\label{fig:transversal}
{An illustration of the transversality property. The logical gate (orange) is  transversal in the sense that it is implemented by physical gates with tensor product forms, i.e.,~acting on individual physical subsystems (yellow). Transversal gates are desirable for fault tolerance because they do not spread errors within code blocks.  In our context, the gates represent symmetry actions so transversality signifies the product property of the symmetry representations on the physical system.}
 }
\end{figure}

A phenomenon that has drawn significant interest lately is that continuous symmetries can induce fundamental limitations on quantum error correction (QEC)\cite{hayden2017error,eastin2009restrictions}, a cornerstone of quantum technologies which was initially introduced as a technique to protect quantum information\cite{shor1995scheme,nielsen2002quantum,gottesman2010introduction,lidar2013quantum} and has since been found to play fundamental roles in many areas in physics including quantum gravity\cite{almheiri2015bulk,pastawski2015holographic} and condensed matter physics\cite{kitaev2003fault,zeng2019quantum,brandao2019quantum,PhysRevLett.120.200503}. An early result is the Eastin--Knill theorem\cite{eastin2009restrictions} which, in a quantum computation language, says that if a (finite-dimensional) quantum code implements any continuous group of gates transversally (see \figref{fig:transversal} for an illustration), it cannot exactly correct local errors. 
It is worth noting that the transversality property is not only important in quantum computation since transversal gates are particularly desirable for fault tolerance, but also widely so in physics as a fundamental feature of internal symmetries in many-body scenarios. 
Further, there has been a series of recent works that consider approximate QEC by covariant codes, providing bounds on the accuracy as well as explicit constructions\cite{faist2019continuous,woods2019continuous,wang2019quasi,kubica2020using,zhou2020new,yang2020covariant,wang2021theory,tajima2021symmetry,KongLiu21:random}.
Viewing the gates as symmetry actions, these results indicate that the QEC accuracy of a code admitting transversal continuous symmetries is necessarily restricted to some extent. 
Such codes are called \emph{covariant codes} in the literature\cite{hayden2017error,faist2019continuous,woods2019continuous}.
Besides the direct relevance to quantum computation, covariant codes have intriguing connections to wide-ranging disciplines in physics including condensed matter physics\cite{brandao2019quantum,PhysRevLett.120.200503,wang2019quasi}, holography\cite{harlow2018symmetries,harlow2018constraints,kohler2019toy,faist2019continuous,woods2019continuous}, and quantum information\cite{hayden2017error,woods2019continuous,KongLiu21:random}.

As certain forms of invariance under transformations, symmetries are by definition exact.  The existing works on QEC with symmetries\cite{hayden2017error,faist2019continuous,woods2019continuous,wang2019quasi,kubica2020using,zhou2020new,yang2020covariant,wang2021theory,tajima2021symmetry,KongLiu21:random} indeed focused on the QEC performance of exactly covariant codes. However,  it is often useful or even necessary to consider approximate forms of symmetries or conservation laws, especially in quantum physics. For instance, noise effects and imperfections that are common in realistic scenarios can cause deviations from the exact symmetry conditions. There are various more fundamental symmetry breaking mechanisms including spontaneous symmetry breaking, anomalies, and non-renormalizable effects\cite{symmetry-breaking}. 
Remarkably, the inexactness of symmetries plays a significant role in our understanding of wide-ranging aspects of fundamental physics. In particle physics, many fundamental symmetries are known to be only approximate\cite{Witten2018}. In fact, it is commonly believed that  global symmetries cannot be exact in a unified theory of quantum mechanics and gravity\cite{Misner1957,Giddings1988,PhysRevD.52.912,Arkani_Hamed_2007,BanksSeiberg11,Witten2018}. Notably, one modern argument justifying the belief in more concrete terms in AdS/CFT is closely related to covariant codes\cite{harlow2018constraints,harlow2018symmetries,faist2019continuous}.
Despite the importance of approximate symmetries, our understanding of them, especially on a quantitative level, is quite limited and unsystematic, raising the need for a general theory of symmetry violation measures. In particular, for QEC and associated problems in physics, it is imperative that we quantitatively understand how symmetry violation is induced by QEC accuracy, which represents a theory of the emergence of approximate symmetries.

In this work, we introduce a variety of approximate continuous symmetry measures in quantum channels and codes, and furthermore, establish a comprehensive theory of the competition between them and QEC accuracy for the most fundamental $U(1)$ case.   
More specifically, we introduce three different meaningful measures of the degree of symmetry violation in terms of group-global and group-local covariance violation respectively, and charge conservation violation, which correspondingly induce different quantitative notions of approximately covariant codes.
We then derive a series of trade-off relations between the QEC inaccuracy and the above approximate symmetry measures under a general condition called Hamiltonian-in-Kraus-span (HKS) condition which subsumes transversality in our setup, by employing various  techniques and ideas from approximate QEC\cite{knill1997theory,beny2010general}, quantum metrology\cite{giovannetti2011advances,kolodynski2013efficient,zhou2020theory} and quantum resource theory\cite{chitambar2019quantum,marvian2020coherence,FangLiu19:nogo,fang2020no}.
In particular, 
 our results indicate various forms of lower bounds on the symmetry violation of QEC codes, which imply universal fine-grained restrictions on the transversally implementable gates, improving the Eastin--Knill theorem. 
To exemplify the general theory, we present two explicit families of approximately covariant codes that nearly saturate certain lower bounds. In the end, we provide a blueprint for several potential applications in quantum gravity and condensed matter physics.

The main goal of this paper is to elucidate the intuitions behind our approaches and report the most representative results.  We refer interested readers to the Supplementary Information~\cite{SM}, which can be read independently as an extended companion paper gearing more towards the technical audience.  All technical details, additional results, and more extensive discussions can be found in it.

\section*{Results}

\subsection{Quantitatively~characterizing~approximate~symmetries.}

We first introduce, from a general standpoint, three physically motivated types of quantitative measures of the symmetry violation of quantum systems.  

Let $G$ be a compact Lie group corresponding to the continuous symmetry of interest.  Denote by $\mE_\AtoB$ a quantum channel from system $A$ to system $B$.   
The channel exactly respects symmetry $G$ if it is \emph{covariant} with respect to the group actions, i.e., 
\begin{equation}
\mE_\AtoB\circ \mU_{A,g} = \mU_{B,g}\circ \mE_\AtoB,
\end{equation}
or equivalently $\mU_{B,g}^\dagger \circ \mE_\AtoB \circ \mU_{A,g} = \mE_\AtoB$, for all $g\in G$, 
where we use $\mU(\cdot) := U(\cdot )U^\dagger$ to denote the channel action of unitary $U$, and $U_g$ is determined by the unitary representation of $G$   under consideration (on the appropriate system).  

To characterize the deviation from the exact symmetry, it is natural to consider the mismatch between the two sides of the covariance condition.
Then an intuitive overall measure is the maximum mismatch overall the entire group as given by some channel distance $D$, which we call \emph{group-global covariance violation}:
\begin{align}
 \tilde\delta_\group := \max_{g\in G} D(\mE_\AtoB\circ \mU_{A,g}, \mU_{B,g}\circ \mE_\AtoB).
\end{align}
Note that we may not explicitly write down the arguments of the measures when they are unambiguous. 

Another important notion is the symmetry violation around a certain point $g_0$ in the group at which the covariance condition  $\mE_{\AtoB}\circ\mU_{A,g_0} = \mU_{B,g_0}\circ \mE_{\AtoB}$ holds. Here we are interested in the local geometry of the mismatch around this point.   Without loss of generality, we assume $g_0 = \id$, because for an arbitrary $g_0$, we can always redefine the quantum channel to be $\mU_{A,g_0}\circ \mE_{\AtoB}$. 
Let the symmetry actions be parametrized by $\bm{\theta} = \{\theta_k\}\in\mathbb{R}^K$ via some unitary representation $U_g = e^{-i \bm{J} \cdot \bm{\theta}}$ where $\bm{J} = \{J_k\}$ are infinitesimal generators of $G$. For all $k = 1,\ldots,K$, we define the \emph{group-local (point) covariance violation} as
\begin{equation}
\label{eq:point-general}
\tilde\delta_{\point,k}:= \big( 2 \partial_{\theta_k}^2 D(\mE_\AtoB\circ \mU_{A,e^{-iJ_k\theta_k}}, \\ \mU_{B,e^{-iJ_k\theta_k}}\circ \mE_\AtoB)^2\big)^{1/2}\big|_{\theta_k=0},
\end{equation}
where $\partial_{\theta}^2 D^2$ denotes the second-order derivative of $D^2$ with respect to $\theta_k$. The square root and the coefficient $\sqrt{2}$ in the definition are chosen to simplify calculations, as will be seen later on. It is worth noting that this notion is closely connected to local parameter estimation, which is a standard setup in metrology, as will be further discussed.

Lastly, it is natural to consider the deviation from conservation laws. Specifically, each generator $J_k$ is associated with a charge, and we can quantify the variation of the charge for input state $\rho$ by 
\begin{equation}
\tilde\delta_{\charge,k}(\rho) :=  \abs{\trace J_{k,B}\mE_\AtoB(\rho) - \trace J_{k,A}\rho},
\end{equation}
where $J_{k,A}$ and $J_{k,B}$ are the generators on systems $A$ and $B$, respectively, so the trace gives the expectation value of the associated charge.  Then overall measures can be defined based on e.g.~maximization over $\rho$.
Note that $\tilde\delta_\charge$ is  not necessarily zero for covariant encoding channel $\mE$, except when $\mE$ is isometric, which is the standard in QEC scenarios\cite{cirstoiu2020robustness}. 

Since $\tilde\delta_\point$ and $\tilde\delta_\charge$ only depend on the local geometry of the symmetry group, we shall collectively call them \emph{local symmetry measures}.  As will become evident, these three measures are inequivalent and their behaviors should be understood independently; notably, $\tilde\delta_\group$ is a unitless measure with range $[0,1]$, in contrast to $\tilde\delta_\point$ and $\tilde\delta_\charge$.

\subsection{QEC~setting: code accuracy and symmetry.}

Now we describe the QEC setup and in particular, formally define the QEC (in)accuracy and  symmetry measures associated with a quantum code (as an application of the general measures introduced above) that will be considered.

In this work, we mainly consider a type of distance measure based on the \emph{purified distance}, which is particularly well behaved and of broad importance in quantum information.
More specifically, the state purified distance is given by \begin{equation}
P(\rho,\sigma) := \sqrt{1 - f(\rho,\sigma)^2}     
\end{equation}
where $f(\rho,\sigma) := \trace(\sqrt{\rho^{1/2}\sigma\rho^{1/2}})$ is the fidelity, based on which one can define the \emph{channel purified distance}\cite{gilchrist2005distance,tomamichel2015quantum,LiuWinter19} as
\begin{equation}
    P(\Phi_1,\Phi_2) := \max_{\rho} P((\Phi_1\otimes\id)(\rho),(\Phi_2\otimes\id)(\rho)).
\end{equation}
In principle, one may also consider other metrics. In Secs.~II and VII of Ref.~\cite{SM}, we also discuss the situations where one uses e.g.~the diamond norm distance\cite{watrous2018theory}, which is another standard channel distance measure. 
Additionally, the purified distance is equal to the Bures distance defined by $\sqrt{2(1 - f(\rho,\sigma))}$ up to the leading order when $f \approx 1$ or the distance is small, a quantity used to study approximate QEC before\cite{beny2010general}. 
The choice of purified distance directly relates the local covariance violation to the quantum Fisher information (QFI)\cite{helstrom1976quantum,holevo1982probabilistic,hubner1992explicit,sommers2003bures}, which is a central notion in quantum metrology as a standard quantifier of the amount of information a parametrized quantum state $\rho_\theta$ carries about the parameter $\theta$ locally. Note that the QFI we use here is conventionally called the symmetric logarithmic derivative (SLD) QFI, while there are other types of QFIs, e.g., the right logarithmic derivative (RLD) QFI\cite{yuen1973multiple,hayashi2011comparison,katariya2020geometric} which we will encounter in Methods. 

A quantum code is defined by an encoding channel $\mE_{\LtoS}$, which is a map from a logical system $L$ to a physical system $S$. The physical system may be subject to a noise channel $\mN_{S}$.  Ideally, if we can find a recovery (decoding) channel $\mR_{\StoL}$ that achieves 
\begin{equation}
\label{eq:qec_identity}
\mR_{\StoL}\circ \mN_{S}\circ\mE_{\LtoS} = \id_L,     
\end{equation}
where $\id_L$ denotes the logical identity channel, indicating that the QEC procedure essentially removes the noise effects so that the logical information is perfectly recovered, we say the code achieves exact QEC or is an QEC code. For general codes, there may not exist a recovery channel that satisfies \eqref{eq:qec_identity}, meaning that QEC cannot be done perfectly.
Naturally, 
 we characterize the \emph{QEC inaccuracy} by
\begin{align}
\epsilon &:= \min_{\mR_{\StoL}} P(\mR_{\StoL}\circ \mN_{S}\circ\mE_{\LtoS}, \id_L), 
\end{align}
that quantifies the minimum deviation from perfect information recovery. 

We now consider symmetry in the QEC setting. Here the encoding $\mE_{\LtoS}$ is the channel of interest in the last section, and the symmetry acts on its input and output systems, namely the logical system $L$ and the physical system $S$.
From here on, we shall base our discussion on $U(1)$ symmetry, which is of fundamental importance in itself and nicely reveals the key phenomena. Specifically, consider the family of logical gates $U_{L,\theta} = e^{-iH_L\theta}$ for every $\theta \in \bR$, implemented by physical gates $U_{S,\theta} = e^{-iH_S\theta}$ for every $\theta \in \bR$, which are $U(1)$ representations on $L$ and $S$ generated by non-trivial Hamiltonians (symmetry generators) $H_L$ and $H_S$, respectively (note that they should not be confused with the intrinsic Hamiltonians of the systems).

Suppose the Kraus decomposition of the noise channel is given by $\mN_{S}(\cdot) = \sum_i K_{S,i}(\cdot)K_{S,i}^\dagger$ where $K_{S,i}$ are the Kraus operators. A sufficient condition for the non-existence of exactly covariant QEC codes, or in other words, the incompatibility between exact symmetry and QEC,  is 
\begin{equation}
    H_S \in {\rm span}\{K_{S,i}^\dagger K_{S,j},\,\forall i,j\}, 
\end{equation}
which will be referred to as the `Hamiltonian-in-Kraus-span' (HKS) condition\cite{zhou2020new}. Note that although the HKS condition is defined using Kraus operators, it is just a function of the noise channel, because different Kraus representations of a channel will give the same Kraus span.
An important scenario obeying the HKS condition that can be regarded a prototypical case of our theory is when $H_S$ is 1-local and $\mN_S$ is any single-qubit noise with a full Kraus span. This includes erasure, depolarizing, and amplitude damping noises\cite{demkowicz2012elusive}.
The 1-local property means that each term in $H_S$ acts nontrivially only on one subsystem and corresponds to the transversality of  $U_{S,\theta}$  (illustrated in \figref{fig:transversal}). As motivated earlier, this transversality property is crucial to practical quantum computing since transversal gates are fault-tolerant. Moreover, it is broadly important in physical contexts, where one normally considers global internal symmetries generated by sums of disjoint local charge observables, in particular on-site (transversal with respect to sites) symmetries; a typical example is  $U(1)$ generated by the Hamming-weight-type charge observable.  Note that whether the symmetries are on-site is linked to whether they can be gauged or are anomaly-free, which plays important roles in the physics of quantum many-body systems and field theories\cite{Wen13}. Transversality also plays fundamental roles in AdS/CFT\cite{harlow2018symmetries,PhysRevLett.117.021601,MaySorceYoshida}.
It is worth noting that, when the HKS condition fails, examples of exactly covariant QEC codes exist, for instance, when $\mN_{S} = \id$ (unitary dynamics), when $H_S$ is a Pauli-X operator and $\mN_S$ is a dephasing noise\cite{kessler2014quantum,arrad2014increasing}, and when $H_S$ is 2-local and $\mN_S$ is a single-erasure noise\cite{gottesman2016quantum}. 
We will assume the HKS condition holds from now on and study the consequent competition between QEC and covariance. 

Finally, we formally define the three quantitative measures of code symmetry (violation) that we shall consider, obtained by applying the general ideas from the last section to our setup where the symmetry group is $U(1)$.
\begin{enumerate}[i)]
\item \emph{Group-global covariance violation}: Directly following the general definition, we consider 
\begin{align}
\delta_\group := \max_\theta P(\mU_{S,\theta}\circ\mE_{\LtoS}, \mE_{\LtoS}\circ\mU_{L,\theta}). 
\end{align}
$\delta_\group = 0$ if and only if the code is exactly covariant. 
\item \emph{Group-local covariance violation}, in the vicinity of a certain $\theta_0$ where the code is exactly covariant (as explained, we assume $\theta_0 = 0$ without loss of generality): 
For our setting, by letting $D$ be the channel purified distance in $\tilde\delta_\point$ we obtain
\begin{equation}
\delta_\point := \sqrt{F(\mU_{S,\theta}\circ\mE_{\LtoS}\circ\mU_{L,\theta}^\dagger)|_{\theta = 0}},  
\end{equation}
where $F(\cdot)$ is the channel QFI given by $F(\Phi_\theta) = \max_{\rho} F((\Phi_\theta\otimes\id)(\rho))$\cite{fujiwara2008fibre}, in which
\begin{equation}
    F(\rho_{\theta}) := 2\frac{\partial^2 P(\rho_{\theta},\rho_{\theta'})^2}{\partial \theta'^2}\Big|_{\theta' = \theta}, 
\end{equation}
is the conventional state QFI\cite{helstrom1976quantum,holevo1982probabilistic,hubner1992explicit,braunstein1994statistical,sommers2003bures}.
(Note that it is standard to take the infinitesimal form of a distance function to define a measure of local sensitivity. Different definitions of QFI will be induced by different distance functions, e.g., the Wigner-Yanase skew information is induced by the Hellinger distance\cite{luo2004informational}. However, we will focus only on the purified distance and the above definition of group-local covariance violation in this work.) 
%This explicitly demonstrates the claimed connection between the group-local covariance violation and QFIs which characterize the possible precision of local parameter estimation.

\item \emph{Charge conservation violation}: Recall that the logical and physical charge observables are, Hermitian operators $H_L$ and $H_S$ respectively.
As mentioned, isometric covariant channels are always charge-conserving, i.e.,~satisfying $H_L - \nu \id = (\mE_{\LtoS})^\dagger(H_S)$ (up to some constant offset $\nu$ which does not affect the $U(1)$ group representations) where $(\mE_{\LtoS})^\dagger$ is the dual of the encoding channel\cite{faist2019continuous} that maps from $S$ to $L$ (the subscript $\StoL$ is omitted for simplicity of notation).
Here we directly consider
\begin{align}
\delta_\charge := \Delta\big(H_L - (\mE_{\LtoS})^\dagger(H_S)\big), 
\end{align}
where $\Delta(H)$ denotes the difference between the maximum and minimum eigenvalues of a Hermitian operator $H$, or equivalently,
\begin{equation}
   \Delta(H) = 2 \max_{\nu \in \bR} \norm{H - \nu\id}.
\end{equation}
(For simplicity of notation $\Delta H$ is used interchageably with $\Delta(H)$.) $\delta_\charge$ is a close variant of the general  $\tilde\delta_\charge$. It can be verified that 
$\delta_\charge = 2\min_{\nu\in\bR} \max_\rho| \trace(H_S\mE_{\LtoS}(\rho))-\trace ((H_L-\nu \id)\rho)|$, 
where we allow a constant offset on the definitions of charges. 
\end{enumerate}

In what follows, we will refer to `group-global' and `group-local' as `global' and `local'  respectively for simplicity as is common in e.g.~estimation theory, which should be distinguished from the geometric notions commonly used in physical contexts.  The remarks on the general measures at the end of the last section still apply here.

\subsection{Symmetry~vs.~QEC.} 

\begin{figure}[H]
\centering
\includegraphics[width=0.7\textwidth]{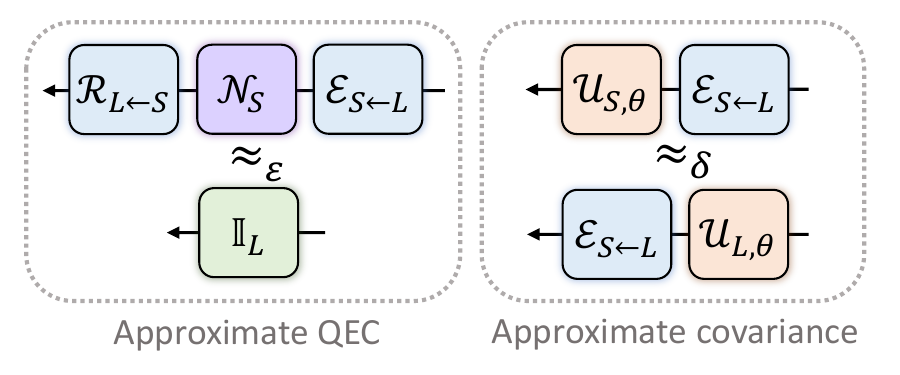}
 \caption{\label{fig:setting}
{Measuring approximate QEC and approximate symmetry. We study the trade-off between QEC inaccuracy (the deviation of the QEC procedure from the logical identity channel, as shown in the left panel) and symmetry violation (the deviation of the encoding map from being covariant with respect to symmetry actions, as shown in the right panel).}
 }
\end{figure}

We now present our main results---general joint bounds of the QEC accuracy and approximate symmetry measures of a code---which reveal fundamental trade-offs between them under the HKS condition (see \figref{fig:setting} for an illustration).  Detailed proofs and further results can be found in Secs.~III and IV of Ref.~\cite{SM}.  When expressing the results, by `$x \gtrsim y$' we mean $x \geq \ell(y)$ for some $\ell(y)$ that is equivalent to $y$ asymptotically (i.e., $\lim_{y\rightarrow 0^+} \ell(y) / y = 1$).

We first discuss the global symmetry violation $\delta_\group$. Using different approaches that will be explained, we prove the following two results: 
\begin{theorem}
\label{thm:global-1}
When $\mE_{\LtoS}$ is isometric,    
\begin{equation}
    \delta_\group \gtrsim \sqrt{\frac{\Delta H_L - 2\epsilon \frakJ(\mN_S,H_S)}{\Delta H_S}},
\end{equation}
where $
\frakJ(\mN_S,H_S) := \min_{h: H_S = \sum_{ij} h_{ij} K_{S,i}^\dagger K_{S,j}} \Delta h$, $h$ is an arbitrary Hermitian matrix. 
\end{theorem}

\begin{theorem}
\label{thm:global-2}
\begin{equation}
    \epsilon + \delta_\group \gtrsim \frac{\Delta H_L}{\sqrt{4\frakF(\mN_S,H_S)}}, 
\end{equation}
where $\frakF(\mN_S,H_S) := 4\min_{h: H_S = \sum_{ij} h_{ij} K_{S,i}^\dagger K_{S,j}} \big\|\sum_{ij}(h^2)_{ij} \allowbreak K_{S,i}^\dagger K_{S,j} - H_S^2\big\|$, $h$ is an arbitrary Hermitian matrix. 
\end{theorem}

\thmref{thm:global-1} and \thmref{thm:global-2} both demonstrate the robust trade-off between $\epsilon$ and $\delta_\group$; specifically, they give non-trivial lower bounds on $\epsilon$ for sufficiently small $\delta_\group$ and vice versa. 
Here, both $\frakJ$ and $\frakF$ are positive functions of the noise channel $\mN_S$ and the physical charge Hamiltonian $H_S$ (different Kraus representations of a noise channel do not induce different values of $\frakJ$ and $\frakF$.).   Note that the definition of $\frakJ$ has a close connection to the HKS condition; $\frakF$ equals the regularized channel QFI of the noisy physical gate $\mN_{S}\circ\mU_{S,\theta}$ (the regularized QFI of a channel $\Phi_\theta$ is defined by $\barF(\Phi_\theta) = \lim_{N\rightarrow\infty} F(\Phi_\theta^{\otimes N})/N$\cite{kolodynski2013efficient,zhou2020theory}). 
The above results can be broadly applied to different noise and charge settings simply by analyzing $\frakJ$ and $\frakF$, which
are efficiently computable using semidefinite programming (see details Secs.~III and IV of Ref.~\cite{SM}).

To be more concrete, we now discuss the specific scaling behaviors of the above general bounds (see \tableref{table}) .
Consider quantum codes on an $n$-partite system with 1-local $H_S$ corresponding to transversal symmetry action (such that $\Delta H_S = O(n)$ and $\Delta H_L = O(1)$). Consider the following important practical noise models:
i)~single-erasure noise, defined by $\mN_S(\cdot) = \frac{1}{n}\sum_{l=1}^n\mN_{S}^{(l)}$ where $\mN_{S}^{(l)}(\cdot) = \ket{\emptyset}\bra{\emptyset}_{S_l} \otimes \trace_{S_l}(\cdot)$ represents the complete erasure on the $l$-th subsystem with other subsystems unaffected (we use $\ket{\emptyset}$ to denote the vacuum state). In this case we have $\frakJ = O(n)$ and $\frakF = O(n^2)$. Then both \thmref{thm:global-1} and \thmref{thm:global-2} give a $\Omega(1/n)$ lower bound on $\epsilon$ for sufficiently small $\delta_\group$; but  for sufficiently small $\epsilon$, \thmref{thm:global-1} gives a $\Omega(1/\sqrt{n})$ lower bound on $\delta_\group$, which is tighter than $\Omega(1/n)$ given by \thmref{thm:global-2}. 
ii)~i.i.d.~erasure noise, defined by $\mN_S(\cdot) = \bigotimes_{l=1}^n\mN_{S_l}$ where $\mN_{S_l}(\cdot) = (1-p_{\rm e})(\cdot)_{S_l} + p_{\rm e} \ket{\emptyset}\bra{\emptyset}_{S_l}\trace_{S_l}(\cdot)$ represents a local erasure on the $l$-th subsystem and $p_{\rm e}$ is the noise rate. In this case we have $\frakJ = O(n)$ and $\frakF = O(n)$. Then both \thmref{thm:global-1} and \thmref{thm:global-2} give a $\Omega(1/\sqrt{n})$ lower bound on $\delta_\group$ for sufficiently small $\epsilon$; but for sufficiently small $\delta_\group$, \thmref{thm:global-2} gives a $\Omega(1/\sqrt{n})$  lower bound on $\epsilon$, which is tighter than $\Omega(1/{n})$ given by \thmref{thm:global-1}.
To sum up, we see that our two bounds can behave differently and complement each other in different important settings. Note that, these lower bounds grow inverse-polynomially with $n$ while the system dimension is exponentially large. That means errors in systems can easily become intolerable in a linear or a  square-root size of circuits of $n$ qudits, demonstrating the strong competition between QEC and continuous symmetries. %For example, without symmetry constraint, random circuits can generate good QEC codes with an exponentially small QEC inaccuracy, while $U(1)$-symmetric random circuits can only generate QEC codes with at most a polynomially small QEC inaccuracy~\cite{Brown13shortrandom,KongLiu21:random}. 
The above discussions can be extended to general noise models, e.g., single-depolarizing noise acts on each subsystem with different probabilities, and the exact values of the lower bounds can be analytically calculated (see discussions in Sec.~III of Ref.~\cite{SM}).

We now explain the main ideas behind the derivation of \thmref{thm:global-1} and \thmref{thm:global-2}. The full proofs can be found in Secs.~III and IV of Ref.~\cite{SM}. In particular, each of the two results builds upon a meaningful quantity we introduce associated with a code and the symmetry, which may be of independent interest.  

\thmref{thm:global-1} comes from a notion that we call \emph{charge fluctuation}, defined by
\begin{equation}
    \chi := \braket{0_L|(\mE_{\LtoS})^\dagger(H_S)|0_L} - \braket{1_L|(\mE_{\LtoS})^\dagger(H_S)|1_L},
\end{equation} 
where $\ket{0_L}$ and $\ket{1_L}$ are eigenstates of $H_L$ corresponding to the largest and the smallest eigenvalues, respectively. For intuition, consider the following extremes. For  exact QEC codes, it is straightforward to see that $\chi = 0$: the Knill--Laflamme conditions\cite{knill1997theory}  indicate that $\Pi K_{S,i}^\dagger K_{S,j} \Pi \propto \Pi$ for all $i,j$ where $\Pi$ is the projection onto the code subspace, so $\bra{0_L}(\mE_{\LtoS})^\dagger(K_{S,i}^\dagger K_{S,j})\ket{0_L} =
\bra{1_L}(\mE_{\LtoS})^\dagger(K_{S,i}^\dagger K_{S,j})\ket{1_L}$; then since $H_S$ can be written as a linear combination of $K_{S,i}^\dagger K_{S,j}$ due to the HKS condition, we have $\bra{0_L}(\mE_{\LtoS})^\dagger(H_S)\ket{0_L} =
\bra{1_L}(\mE_{\LtoS})^\dagger(H_S)\ket{1_L}$. That is, the value of $\chi$ characterizes the degree of deviation from Knill--Laflamme conditions. 
On the other hand, for exactly covariant codes, we have $\chi = \Delta H_L$ because $(\mE_{\LtoS})^\dagger(H_S) = H_L - \nu \id$ for some $\nu \in \bR$\cite{faist2019continuous}. 
More generally, for approximately QEC or covariant codes,  $\epsilon$ (or $\delta_\group$) can be lower bounded using the distance between $\chi$ and $0$ (or $\Delta H_L$). 
Indeed, \thmref{thm:global-1} is established by combining the following two inequalities for isometric encoding channels, 
\begin{gather}
    \epsilon \geq \abs{\chi}/(2\frakJ) ,\label{eq:chi-epsilon}\\
    \delta_\group \gtrsim \sqrt{\abs{\Delta H_L - \chi}/\Delta H_S},\label{eq:chi-delta}
\end{gather}
where \eqref{eq:chi-epsilon} is derived from the approximate Knill--Laflamme conditions\cite{beny2010general}, and \eqref{eq:chi-delta} follows from the definition of $\delta_\group$.  

\thmref{thm:global-2} is derived from another notion that we call \emph{gate implementation error}, defined by 
\begin{equation}
    \gamma := \min_{\mR_{\StoL}} \max_\theta P(\mR_{\StoL}\circ \mN_{S}\circ\mU_{S,\theta}\circ\mE_{\LtoS}, \mU_{L,\theta}), 
\end{equation}
which quantifies the error of implementing an ideal set of logical gates $\mU_{L,\theta}$ using the error-corrected noisy gates $\mR_{\StoL}\circ\mN_{S}\circ\mU_{S,\theta}\circ\mE_{\LtoS}$. The gate implementation error unifies QEC accuracy and symmetry in a sense: it can be proven that
\begin{equation}
    \delta_\group + \epsilon \geq \gamma, 
\end{equation}
putting $\epsilon$ and $\delta_\group$ on the same footing. 
A crucial observation here is that these quantities can be understood from a quantum metrology (channel estimation) perspective: intuitively, smaller error goes hand in hand with higher sensitivity of parameter estimation under noise. This allows us to make use of  quantum metrology and QFI techniques to analyze $\gamma$. Specifically, we show that
\begin{equation}
\label{eq:metrology-1}
    \frakF \gtrsim (\Delta H_L)^2/(4\gamma^2), 
\end{equation}
which implies \thmref{thm:global-2}.
We include in the Methods section further  explanations and details of the quantum metrology method.

Furthermore, using quantum resource theory methods, we can derive different versions of \thmref{thm:global-2} which, in particular, give results on the \emph{average-case} behavior over different input states, in addition to the worst-case results discussed above. Again, see Methods and Sec.~IV of Ref.~\cite{SM} for further explanations and details of the quantum resource theory method.

For the local symmetry measures $\delta_\point$ and $\delta_\charge$, we establish the following trade-off bounds which are also expressed in terms of the general, efficiently computable quantities $\frakJ$ and $\frakF$ that encode the noise and charge Hamiltonian:
\begin{theorem}
\label{thm:charge}
When $\mE_{\LtoS}$ is isometric, 
\begin{equation}
    \epsilon \geq \frac{\Delta H_L - \delta_\charge}{2\frakJ(\mN_S,H_S)} \geq \frac{\Delta H_L - \delta_\point}{2\frakJ(\mN_S,H_S)},
\end{equation}
where $
\frakJ(\mN_S,H_S) := \min_{h: H_S = \sum_{ij} h_{ij} K_{S,i}^\dagger K_{S,j}} \Delta h$, $h$ is an arbitrary Hermitian matrix.
\end{theorem}
\begin{theorem}
\label{thm:local}
    \begin{equation}
        \label{eq:metrology-2}
        \epsilon \gtrsim \frac{\Delta H_L - \delta_\point}{\sqrt{4\frakF(\mN_S,H_S)}},
    \end{equation} 
    where $\frakF(\mN_S,H_S) := 4\min_{h: H_S = \sum_{ij} h_{ij} K_{S,i}^\dagger K_{S,j}} \big\|\sum_{ij}(h^2)_{ij} \allowbreak K_{S,i}^\dagger K_{S,j} - H_S^2\big\|$, $h$ is an arbitrary Hermitian matrix. 
\end{theorem}

Again, \thmref{thm:charge} and \thmref{thm:local} indicate trade-offs between $\epsilon$ and $\delta_\point$ (or $\delta_\charge$), and set non-trivial lower bounds on $\epsilon$ for sufficiently small $\delta_\point$ (or $\delta_\charge$) and vice versa. The proof of \thmref{thm:charge} uses \eqref{eq:chi-epsilon} and the following inequalities for isometric encoding channels:
\begin{equation}
    \abs{\chi} \geq \Delta H_L - \delta_\charge, 
\end{equation}
which follows from the definitions, and a simple relation between $\delta_\point$ and $\delta_\charge$,
\begin{equation}
    \delta_\point \geq \delta_\charge. 
\end{equation}
Similar to \thmref{thm:global-2}, \thmref{thm:local} is derived using quantum metrology for quantum channel techniques. See also Methods and Sec.~VI of Ref.~\cite{SM} for explanations and details. 

Note that $\delta_\point$ and $\delta_\charge$ have the same units as the Hamiltonians and may naturally be superconstant (in contrast to $\delta_\group$ which is always no larger than one). For example, consider the trivial encoding where $L=S$ and $\mE_{\LtoS} = \id$. Then we have $\delta_\point=\delta_\charge=\Delta(H_S-H_L) = \Theta(n)$ for an $n$-partite system with 1-local $H_S$ such that $\Delta H_S = O(n)$ and $\Delta H_L = O(1)$, implying that  a constant scaling of $\delta_\point$ or $\delta_\charge$ still requires non-trivial code structures.

Discussions on further refinements of \thmref{thm:global-1} and \thmref{thm:charge} using quantum metrology techniques, as well as remarks on non-compact groups and infinite-dimensional codes, can be found in Appx.~D and Sec.~II of Ref.~\cite{SM}.

When setting $\delta_\group = \delta_\point = 0$, our theorems recover previous results on exactly covariant codes\cite{faist2019continuous,woods2019continuous,kubica2020using,zhou2020new,yang2020covariant}.  The results and methods here apply to  general quantum codes beyond exactly covariant codes.  In particular, they enable us to quantitatively understand the symmetry restrictions on  exact QEC codes which are  of utmost interest.
In the following section, we demonstrate a particularly important application of our symmetry bounds for QEC codes.

\subsection{General~limitations~on~transversal~gates.}
As motivated earlier, transversal gates or symmetry actions are of central importance to fault-tolerant quantum computing and also widely important in physical contexts. 
Recall that our theory indicates non-trivial bounds for symmetry under the HKS condition, like transversal symmetry. 
Remarkably, by analyzing such bounds, we are able to prove the following fundamental restriction on the set of transversal logical gates for general QEC codes, which refines the Eastin--Knill theorem: 
\begin{corollary}
\label{col:gate}
Suppose an $n$-qudit QEC code with distance at least $2$ admits a transversal implementation $V_S = \bigotimes_{l=1}^n e^{-i2\pi T_{S_l}/D}$ of the logical gate $V_L = e^{-i2\pi T_L/D}$ where $D$ is a positive integer and $T_{L,S}$ are Hermitian operators with integer eigenvalues.
Then $D$ is at most  
\begin{equation}
    O\bigg(\max\bigg\{\frac{(\sum_{l=1}^n (\Delta T_{S_l})^{3/2}}{\sqrt{\Delta T_L}},\Delta T_L\bigg\}\bigg). 
\end{equation}
\end{corollary}
Intuitively, $D$ characterizes the precision or density of the gates.
A key insight behind  \corollaryref{col:gate} is that in the $D\rightarrow \infty$ limit, the code becomes arbitrarily close to an exactly covariant code, which is in conflict with the incompatibility between QEC and continuous symmetries. Further, we obtain the upper bound on $D$ by transforming the lower bound on $\delta_\group$ obtained by applying \thmref{thm:global-1} to exact QEC codes (setting $\epsilon = 0$) with $H_S = \sum_{l=1}^n T_{S_l}$ and $H_L = T_L$ (see Sec.~V of Ref.~\cite{SM}). 

As a standard example, consider the Pauli-Z rotation corresponding to $T_L= Z_L$, $T_{S_l} = - Z_l$; that is, $D$ characterizes the precision of the Pauli-Z rotation. Our \corollaryref{col:gate} gives the upper bound $D = O(n^{3/2})$. On the achievability side, there exists a set of $[[n=2^t-1,1,3]]$ ($t\geq 3$) quantum Reed--Muller codes\cite{steane1999quantum} (which will be further discussed later) 
such that
the logical Pauli-Z rotation $V_L = e^{-i\pi Z_L/2^{t-1}}$ is implemented by transversal physical gate $V_S = \bigotimes_{l=1}^n e^{i\pi Z_l/2^{t-1}}$, i.e.,~achieves $D = 2^t = \Omega(n)$, which is polynomially close to our upper bound $O(n^{3/2})$.

For stabilizer codes, \corollaryref{col:gate} implies that $\tilde{V}_S = \bigotimes_{l=1}^n e^{-i 2\pi a_l Z_l/D}$ where $a_l$ is an integer and $D$ is a power of two (which is the most general form of transversal logical gates up to local Clifford equivalences\cite{zeng2011transversality,anderson2016classification}) can only implement logical gates up to the \mbox{$O(\log n)$-th} level of the Clifford hierarchy when $a_l = O(\poly(n))$ (see Sec.~V of Ref.~\cite{SM}), which can be attained by the aforementioned quantum Reed--Muller codes.
Note that several similar bounds\cite{zeng2011transversality,bravyi2013classification,pastawski2015fault,anderson2016classification,jochym2018disjointness} were known for stabilizer codes. A key point of our results is that such restrictions fundamentally stem from symmetry principles and hold generally for arbitrary codes, meaning that the stabilizer structure is not essential here. 
We remark that there are ways to circumvent the Eastin--Knill theorem and the above limitations that involves more complex QEC procedures such as code switching\cite{anderson2014fault} and magic state injection\cite{gottesman1999demonstrating,bravyi2005universal}.

\subsection{Concrete code examples.}

We now discuss two specific families of quantum codes that exhibit important approximate covariance features and also provide evidence of the tightness of our bounds (more details in Sec.~VII of Ref.~\cite{SM}).

The first example we give is a family of exact QEC codes that are approximately covariant.  
Specifically, let us consider the $[[n = 2^t -1,1,3]]$ ($t \geq 3$) quantum Reed--Muller code\cite{steane1999quantum}. The codewords are 
\begin{align}
\ket{\frakc_0} & 
= \frac{1}{\sqrt{2^t}} \bigg(\ket{\v{0}} + \sum_{\v{x} \in \overline{R(1,t)}\backslash\{\v{0}\}} \ket{\v{x}}\bigg), \\
\ket{\frakc_1} &
= \frac{1}{\sqrt{2^t}} \bigg(\ket{\v{1}} + \sum_{\v{x} \in \overline{R(1,t)}\backslash\{\v{0}\}} \ket{\v{1} + \v{x}}\bigg),
\end{align}
where we use $\v{x}$ to denote $n$-bit strings (${\v{0}}$ and ${\v{1}}$ are all zero and all one strings, respectively) and $\overline{R(1,t)}$ is the classical shortened Reed--Muller code\cite{macwilliams1977theory}. Note that all strings in $\overline{R(1,t)}\backslash\{\v{0}\}$ have weight $2^{t-1}$.
Consider the single-erasure noise which is exactly correctable, namely, $\epsilon = 0$. 
As mentioned above, the code admits a transversal implementation $\bigotimes_l \big(e^{i\pi Z_l/2^{t-1}}\big)$ of the logical operator $e^{-i\pi Z_L/2^{t-1}}$ (here the symmetry is defined by $H_L = \frac{1}{2}Z_L$, $H_S = -\frac{1}{2}\sum_{l=1}^n Z_l$). Intuitively, the larger $t$ or $n$ (associated with the precision of transversal gates) is, the closer the code is to being covariant. Indeed, calculations indicate that $\delta_\group \simeq \sqrt{4/n}$ for large $n$ (`$\simeq$' indicates equivalence at the leading order), saturating our lower bound $\delta_\group \gtrsim \sqrt{1/n}$ up to a constant factor. We see that $\delta_\group$ nicely captures the relation between global gate precision and the closeness to covariance, leading to results like \corollaryref{col:gate}.  As for the local symmetry measures, we find that here $\delta_\point = \sqrt{n+1}$ for all $\theta = \frac{4k\pi}{n+1}$ ($\forall k\in \bZ$) at which the code is exactly covariant, having a gap with its constant lower bound $\delta_\point \gtrsim 1$, and $\delta_\charge = 1$, exactly matching the lower bound $\delta_\charge \geq 1$. 

In the second example, we construct a parametrized family of codes that exhibits a smooth transition from exact covariance to exact QEC, which we call modified thermodynamic code, based on the previously studied thermodynamic code\cite{brandao2019quantum,faist2019continuous}. Our modified thermodynamic code is an $n$-qubit $2$-dimensional quantum code defined on a spin chain with Hamiltonian $H_S = - \frac{1}{2}\sum_{l=1}^n Z_l$ given by codewords 
\begin{align}
\ket{\frakc^q_0} &= \sqrt{\frac{n}{n+qm}} \ket{m_n} + \sqrt{\frac{qm}{n+qm}} \ket{(-n)_n}, \\
\ket{\frakc^q_1} &= \sqrt{\frac{n}{n+qm}} \ket{(-m)_n} + \sqrt{\frac{qm}{n+qm}} \ket{n_n},  
\end{align}
where $q$ is a tunable parameter in $[0,1]$ and it is assumed that $m+n$ is an even number with $2\leq m \ll n$, and $\ket{l_n}$ are Dicke states, i.e., symmetric eigenstates of $H_S$, satisfying $H_S \ket{l_n} = \frac{l}{2} \ket{l_n}$. To understand the code, let us now consider different values of $q$.
When $q = 0$, the code reduces to the thermodynamic  code\cite{brandao2019quantum,faist2019continuous}, which is exactly covariant with respect to $H_L = \frac{m}{2}Z_L$, due to the fact that the codewords are eigenstates of $H_S$ with eigenvalues $\pm \frac{m}{2}$. When $q = 1$, the code satisfies the Knill--Laflamme conditions and is exactly error-correcting under the single-erasure noise. When $q$ is taken in between $0$ and $1$, the code interpolates between the above two extreme cases. Here, we find that $\delta_\group \simeq \sqrt{4qm/n}$ and $\epsilon \simeq (1-q)m/2n$, saturating the scaling of their lower bounds $\delta_\group \gtrsim \sqrt{qm/n}$ and $\epsilon \gtrsim (1-4q)m/2n$, respectively. As for the local symmetry measures, we find that $\delta_\point \simeq \sqrt{qmn}$ for all $\theta = \frac{4k\pi}{n+m}$ ($\forall k\in \bZ$) at which the code is exactly covariant, having a gap with its lower bound $\delta_\point \gtrsim qm$, and $\delta_\charge = qm$, exactly matching the lower bound $\delta_\charge \gtrsim qm$.  In conclusion, both examples achieve the optimal scalings of $\delta_\group$ and $\delta_\charge$.

\subsection{Potential~applications~to~physics.} 

Here we point out a few important areas in physics where our theory is potentially useful.

First, we expect our study to lead to new quantitative insights into the crucial symmetry problem in quantum gravity (see e.g.,\cite{Misner1957,Giddings1988,PhysRevD.52.912,BanksSeiberg11,Arkani_Hamed_2007,harlow2018symmetries,Palti2019}), through the following lenses: 
\begin{enumerate}[i)]
    \item Holography and AdS/CFT correspondence:  AdS/CFT is known to have fundamental connections with QEC\cite{almheiri2015bulk,pastawski2015holographic} and indeed, the no-global-symmetry arguments of Harlow and Ooguri\cite{harlow2018constraints,harlow2018symmetries} is underpinned by insights from QEC. In particular, for the continuous symmetry case, the situation becomes largely equivalent to Eastin--Knill (due to the transversality  deduced from inherent properties of AdS/CFT\cite{almheiri2015bulk,PhysRevLett.117.021601,harlow2018constraints,harlow2018symmetries,faist2019continuous}), 
indicating that our theory can potentially be used to establish  quantitative statements about the breaking of global symmetries in AdS/CFT.  
\item Black hole evaporation:  A standard no-global-symmetry argument is based on certain inconsistencies between the  evaporation and charge conservation of charged black holes\cite{BanksSeiberg11} (note that the weak gravity conjectures\cite{Arkani_Hamed_2007,2022arXiv220108380H} are closely relevant).  Our results may be applied to symmetric versions of Hayden--Preskill model of black hole evaporation as the model can be formulated in terms of QEC (see also \cite{Yoshida:softmode,liu2020,Nakata20,tajima2021symmetry}), through which new insights on charged black hole evaporation may be obtained.
\end{enumerate}
In these scenarios, our theory can potentially be used to derive interesting quantitative bounds on the magnitude of symmetry-violating effects (operators, terms, modes, etc.). 
Note that there are some recent field or string theory calculations on approximate symmetries in quantum gravity\cite{Fichet2020,Hsin21,Chen2021} and it could be intriguing to draw comparisons with our quantum information results.

Furthermore, QEC and symmetries naturally arise together in various key areas in many-body physics like topological phases of matter\cite{Bravyi2010,Michalakis2013,Kitaev2003,Yoshida15:color,Yoshida2017,zeng2019quantum,RobertsBartlett20,wang2021comparative,moessner_moore_2021} and information scrambling\cite{PhysRevX.8.031057,PhysRevX.8.031058,Rakovszky19:renyi,Znidaric2020,Huang20,Yoshida:softmode,tajima2021symmetry,KudlerFlam,KongLiu21:random,Huang22}, where the interplay between them is expected to find interesting applications.  To be more specific, note first that the notions of many-body entanglement,  topological and quantum order, and QEC are intimately connected\cite{zeng2019quantum,Bravyi2010,Kitaev2003,Yoshida15:color,moessner_moore_2021}; besides, as mentioned above, the violation of symmetry conditions is important in realistic systems and in particular, goes hand in hand with anomalies, which play fundamental roles in the physics of quantum systems in various ways.  Our study potentially provides a powerful framework for establishing rigorous connections among these important concepts.

\section*{Discussion}

In this work, we  developed a systematic quantitative theory of the interplay between continuous symmetries and QEC by introducing several notions of approximate symmetry measures based on both global and local symmetry violation and analyzing  QEC accuracy together with them in quantum codes.
A key message is that the degree of symmetry (in multiple senses) and optimal QEC accuracy of a code are concurrently limited by trade-off relations between them, which has interesting implications in quantum computation and physics.

We point out a few directions that are worth further study. {First,
it would be interesting to further understand whether the two trade-off relations between global symmetry and QEC, which exhibit different behaviours under different noise models, can be unified.}
Another natural future task is to extend our study to more general continuous symmetry groups including non-Abelian ones with multiple non-commuting charges, in particular,  $SU(d)$ symmetry, which will complete the  understanding of the limitations on the ability to perform universal quantum computation using transversal gates. More involved representation theory machinery is expected to be useful\cite{faist2019continuous,KongLiu21:random} in the $SU(d)$ extension. 
Discrete symmetries are also important and worth further exploring---although they are not as fundamentally incompatible with QEC as continuous symmetries\cite{hayden2017error}, there exist important scenarios where the incompatibility arises from simple additional constraints (e.g., AdS/CFT codes\cite{harlow2018constraints,harlow2018symmetries,faist2019continuous})---it would be interesting to have a more systematic understanding of the discrete symmetry case (note that this paper readily implies some results). 
Finally, in the last section we pointed to a few potentially relevant physical problems. It would be worthwhile to further consolidate these connections in physics languages, which would enrich the interaction between quantum information and physics and open new doors for both.

\begin{methods}

 We employed methods and techniques native to the fields of quantum metrology and quantum resource theory in our derivation, which we overview here (further details can be found in Secs.~III and IV of Ref.~\cite{SM}).

\subsection{Quantum metrology method.}

First, we describe the quantum metrology method used in the proofs of \thmref{thm:global-2} and \thmref{thm:local}.

As introduced in the main text, we consider the regularized channel QFI\cite{kolodynski2013efficient,zhou2020theory} of the noisy physical gate $\mN_{S,\theta} := \mN_{S}\circ\mU_{S,\theta}$ which is a function of $\mN_S$ and $H_S$ only. 
We have 
\begin{equation}
    \frakF(\mN_S,H_S) = \barF(\mN_{S,\theta}) = \lim_{N\rightarrow \infty} \frac{F(\mN_{S,\theta}^{\otimes N})}{N},
\end{equation}
where $F(\cdot)$ is the channel QFI given by $F(\Phi_\theta) = \max_{\rho} F((\Phi_\theta\otimes\id)(\rho))$\cite{fujiwara2008fibre}.
The goal is to relate it with the error quantities.

The regularized channel QFI possesses the following useful properties: i)~Monotonicity: for arbitrary $\Phi_\theta$ and $\Phi_{1,2}$ independent of $\theta$, it holds that
\begin{equation}
    \barF(\Phi_1 \circ (\Phi_{\theta} \otimes \id) \circ \Phi_2 ) \leq \barF(\Phi_{\theta}).
\end{equation}
It indicates that any type of parameter-independent superchannel, including quantum error correction, cannot increase the value of the regularized channel QFI; 
ii)~$\barF(\mN_{S,\theta}) \geq 0$, and $\barF(\mN_{S,\theta}) = +\infty$ if and only if the HKS condition is violated. As a result, when the HKS condition is satisfied as we assume, there do not exist encoding and decoding channels $\mE_{\LtoS}$ and $\mR_{\StoL}$ such that
\begin{equation}
\label{eq:hks-no-go}
    \mR_{\StoL}\circ\mN_{S,\theta}\circ\mE_{\LtoS} = \mU_{L,\theta},
\end{equation}
because  $\barF(\mU_{L,\theta}) = + \infty$ while $\barF(\mN_{S,\theta}) < +\infty$, so \eqref{eq:hks-no-go} is forbidden by the monotonicity property. 

In order to derive the results, consider an error-corrected noise channel
\begin{equation}
\label{eq:noise-c}
    \mN_{C,\theta} = \mR_{\SAtoC} \circ (\mN_{S,\theta}\otimes \id_A) \circ \mE_{\CtoSA}.
\end{equation}
Here, we introduce an ancillary qubit system $A$ and a logical qubit system $C$, and consider specifically the  encoding and decoding channels 
\begin{gather}
    \mR_{\SAtoC} = \mR^{\rep}_{\LAtoC} \circ (\mR_{\StoL}\otimes \id_A), \\
    \mE_{\CtoSA} = (\mE_{\LtoS}\otimes \id_A)\circ \mE^{\rep}_{\CtoLA},
\end{gather}
where $\mE_{\LtoS}$ is the quantum code under study, $\mE^{\rep}_{\CtoLA}$ is the repetition code $\mE^{\rep}_{\CtoLA}(\ket{i_C})=\ket{i_C i_A}$ for $i=0,1$, and $\mR^{\rep}_{\CtoLA}$ is the decoding channel that perfectly corrects bit-flip errors on $L$. 
We show that for any $\mR_{\StoL}$, $\mN_{C,\theta}$ is a rotated dephasing channel on $C$ satisfying 
\begin{equation}
\label{eq:dephasing_f}
    \mN_{C,\theta}(\ket{i_C}\bra{j_C}) = \delta_{ij}  \ket{i_C}\bra{j_C} + (1 - \delta_{ij}) \xi_\theta \ket{i_C}\bra{j_C},
\end{equation}
for all $i,j=0,1$ and some complex number $\xi_\theta$ (the angle of $\xi_\theta$ indicates the angle of Pauli-Z rotation and the magnitude of $\xi_\theta$ indicates the dephasing rate). 
The regularized channel QFI of the rotated dephasing channel is given by 
\begin{equation}
    \barF(\mN_{C,\theta}) = \frac{\abs{\partial_\theta \xi_\theta}^2}{1 - \abs{\xi_\theta}^2}, 
\end{equation}
where $\abs{\partial_\theta \xi_\theta}$ characterizes the strength of the signal and $\sqrt{1 - \abs{\xi_\theta}^2}$ characterizes the strength of the dephasing noise; that is, $\barF(\mN_{C,\theta})$ can be handwavily understood as the square of the signal-to-noise ratio. By some calculations based on \eqref{eq:dephasing_f}, we obtain (for $\delta_\point < \Delta H_L$)
\begin{gather}
\label{eq:dephasing-1}
    \barF(\mN_{C,\theta}) \gtrsim \frac{(\Delta H_L)^2}{4\gamma^2} \text{~~~for some }\theta, \\
\label{eq:dephasing-2}
    \barF(\mN_{C,\theta}) \gtrsim \frac{(\Delta H_L - \delta_\point)^2}{4\epsilon^2} \text{~~~for some }\theta. 
\end{gather}
Now note that by the monotonicity of the regularized channel QFI, we have 
\begin{equation}
\label{eq:monotonicity}
    \barF(\mN_{C,\theta})  \leq \barF(\mN_{S,\theta}) = \frakF(\mN_S,H_S), ~\forall \theta.
\end{equation}
By combining \eqref{eq:monotonicity} and \eqref{eq:dephasing-1}, we obtain \eqref{eq:metrology-1}, namely $\frakF \gtrsim (\Delta H_L)^2/(4\gamma^2)$,
which implies  \thmref{thm:global-2}; and by  combining \eqref{eq:monotonicity} and \eqref{eq:dephasing-2}, we obtain \eqref{eq:metrology-2} (\thmref{thm:local}), namely $\epsilon \gtrsim {(\Delta H_L - \delta_\point)}/{\sqrt{4\frakF}}$.

\subsection{Quantum resource theory method.}

Here we introduce a different line of thought, which draws ideas and methods from another active field, namely quantum resource theory. More specifically, inspired by the no-purification theories\cite{FangLiu19:nogo,fang2020no,marvian2020coherence}, we can bound the QEC inaccuracy and the global covariance violation jointly using the monotonicity of suitable asymmetry measures.

In particular, while the theorems introduced in the main text are worst-case (with respect to all input states) results, the resource theory method
allows us to obtain average-case results. Specifically, we show that
\begin{equation}
\label{eq:res}
    \bepsilon + \bdelta_\group \gtrsim \sqrt{\frac{\frac{1}{d_L}\trace(H_L^2)-\frac{1}{d_L^2}\trace(H_L)^2}{F^\txr(\mN_{S,\theta})}},
\end{equation}
assuming that $\mN_{S}$ commutes with $\mU_{S,\theta}$. 
Here $\bepsilon$ and $\bdelta_\group$ are, respectively, what we call the Choi QEC inaccuracy and Choi global covariance violation defined by 
\begin{gather}
    \bepsilon := \min_{\mR_{\StoL}} \barP(\mR_{\StoL}\circ \mN_{S}\circ\mE_{\LtoS}, \id_L), \\ 
    \bdelta_\group := \max_\theta \barP(\mU_{S,\theta}\circ\mE_{\LtoS},\mE_{\LtoS}\circ\mU_{L,\theta}),
\end{gather}
where  $\barP(\Phi_1,\Phi_2) := P((\Phi_1\otimes \id)(\Psi),(\Phi_2\otimes \id)(\Psi))$ with the maximally entangled state $\ket{\Psi} = \frac{1}{\sqrt{d}}\sum_{i=1}^d\ket{i}\ket{i}$ ($d$ is the input dimension of $\Phi_{1,2}$) as the input state, is the Choi purified distance. The Choi measures capture the average-case behaviors in the sense that they are closely related to the uniform averages over all pure input states given by integration over the Haar measure\cite{horodecki1999general,nielsen2002simple,gilchrist2005distance}. 
Furthermore, $F^\txr(\mN_{S,\theta})$ is the right logarithmic derivative (RLD) channel QFI\cite{hayashi2011comparison} defined by 
\begin{equation}
    F^\txr(\Phi_\theta) = \max_{\rho} F^\txr((\Phi_\theta\otimes\id)(\rho)),
\end{equation}
where the RLD state QFI\cite{yuen1973multiple} $F^\txr(\rho_\theta)$ is equal to $ \trace((\partial_\theta \rho_\theta)^2\rho_\theta^{-1})$ when $\supp(\partial_\theta \rho_\theta) \subseteq \supp(\rho_\theta)$, and $+\infty$ otherwise.

In order to prove \eqref{eq:res}, we use a resource theory of asymmetry\cite{marvian2020coherence} where the free (incoherent) states are quantum states that commute with the  Hamiltonian and the free (covariant) operations $\mC_{\StoL}$ are quantum operations that commutes with the symmetry actions, namely satisfying 
\begin{equation}
\label{eq:twirl}
    \mC_{\StoL} \circ \mU_{S,\theta} = \mU_{L,\theta} \circ \mC_{\StoL},~\forall \theta. 
\end{equation}
Here the RLD QFI induces a resource monotone: 
\begin{equation}
    F^\txr(\rho,H) := F^\txr(e^{-iH\theta}\rho e^{iH\theta}).
\end{equation}
It can be easily seen that it is infinite for pure coherent states. 

The intuition goes as follows. First, we prove that there always exists a covariant recovery channel $\mR^\cov_{\StoL}$ that can achieve QEC inaccuracy  
\begin{equation}
\label{eq:res-1}
   \bepsilon_\cov = \min_{\mR^\cov_\StoL}
\barP(\mR^\cov_{\StoL}\circ \mN_{S}\circ\mE_{\LtoS}, \id_L)\leq \bepsilon + \bdelta_\group,
\end{equation}
assuming $\mN_S$ commutes with $\mU_{S,\theta}$.  This is done by construction: we show that a  variant of an optimal recovery channel achieving $\bepsilon$, obtained by a suitable twirling action over the symmetry group (which implies that the channel is covariant), satisfies \eqref{eq:twirl}.
The monotonicity of the resource monotone indicates that
\begin{equation}
\label{eq:res-2}
    F^\txr(\mN_S\circ\mE_{\LtoS}(\Psi_{LR}),H_S\otimes \id_R)\geq  F^\txr(\mR_{\StoL}^{\cov}\circ \mN_S\circ\mE_{\LtoS}(\Psi_{LR}),H_L\otimes \id_R),
\end{equation}
where $\ket{\Psi_{LR}} = \frac{1}{\sqrt{d_L}}\sum_{i=1}^{d_L}\ket{i_L}\ket{i_R}$ is a maximally entanglement state between the logical system $L$ and a reference system $R$. The left-hand side is no larger than the RLD channel QFI $F^\txr(\mN_{S,\theta})$ due to the definition of the channel QFI. Notice that the right-hand side tends to infinity when $\mR_{\StoL}^{\cov}\circ \mN_S\circ\mE_{\LtoS}$ tends to the identity channel due to the property that $F^\txr(\rho,H)$ is infinite for pure coherent states. By a more detailed analysis, we can show that
\begin{equation}
\label{eq:res-3}
    F^\txr(\mR_{\StoL}^{\cov}\circ \mN_S\circ\mE_{\LtoS}(\Psi_{LR}),H_L\otimes \id_R)  \geq  \frac{\frac{1}{d_L}\trace(H_L^2)-\frac{1}{d_L^2}\trace(H_L)^2}{\bepsilon_\cov^2},
\end{equation}
which links $\bepsilon_\cov$ with the resource measures. Combining \eqref{eq:res-1}, \eqref{eq:res-2} and \eqref{eq:res-3} , we obtain \eqref{eq:res}. 

Following an analogous argument, a trade-off relation between the worst-case $\epsilon$ and $\delta_\group$ (which turns out to be weaker than \thmref{thm:global-2}; see Sec.~IV of Ref.~\cite{SM} for details) can also be derived using the quantum resource theory method.

\end{methods}

\begin{addendum}
 \item[Data Availability]  There is no data in this paper. 
 \item 
We thank Daniel Gottesman, Liang Jiang, Dong-Sheng Wang, Weicheng Ye, Jinmin Yi, Beni Yoshida for valuable discussions and feedback. 
Z.-W.L. is supported by Perimeter Institute for Theoretical Physics and Yau Mathematical Sciences Center, Tsinghua University.
Research at Perimeter Institute is supported in part by the Government of Canada through the Department of Innovation, Science and Economic Development Canada and by the Province of Ontario through the Ministry of Colleges and Universities. 
S.Z. acknowledges funding provided by the Institute for Quantum Information and Matter, an NSF Physics Frontiers Center (NSF Grant PHY-1733907) and Perimeter Institute for Theoretical Physics. S.Z. also acknowledges support through the University of Chicago from ARO (W911NF-18-1-0020, W911NF-18-1-0212), ARO MURI (W911NF-16-1-0349), AFOSR MURI (FA9550-19-1-0399, FA9550-21-1-0209), DoE Q-NEXT, NSF (EFMA-1640959, OMA-1936118, EEC-1941583), NTT Research, and the Packard Foundation (2013-39273), where part of this work was done. 

 \item[Competing Interests] The authors declare no competing financial or non-financial interests. 
 
 \item[Author Contributions] 

 Z.-W.L. and S.Z. jointly developed the work and wrote the manuscript. 

 %\item[Correspondence] Correspondence and requests for materials should be addressed to Z.-W.L. (email: zwliu0@mail.tsinghua.edu.cn) and S.Z. (email: sisi.zhou26@gmail.com).

\end{addendum}

\vspace{0.4in}
\bibliographystyle{naturemag}
\bibliography{refs-approx-cov-nc}

\newpage

\begin{table}[h]
\begin{center}
\begin{tabular}{|c || c | c | c | c|} 
 \hline
  Lower bounds on  & $\epsilon$ (\thmref{thm:global-1}) & $\epsilon$ (\thmref{thm:global-2}) & $\delta$ (\thmref{thm:global-1}) & $\delta$ (\thmref{thm:global-2}) \\ [0.5ex] 
 \hline\hline
\begin{tabular}{@{}c@{}}Random local noise \\ e.g., $\mN_S = \sum_{l=1}^n \frac{1}{n} \mN_{S_l}$ \end{tabular} 
 %Random local noise ($\mN_S = \sum_{l=1}^n q_l \mN_{S_l}$, $\sum_l q_l = 1)$ 
 & $\Omega\left(\frac{1}{n}\right)$ & $\Omega\left(\frac{1}{n}\right)$ & $\Omega\left(\frac{1}{\sqrt{n}}\right)$ & \textcolor{gray}{$\Omega\left(\frac{1}{n}\right)$} \\ 
 \hline
 \begin{tabular}{@{}c@{}} Independent noise \\ e.g., $\mN_S = \bigotimes_{l=1}^n \mN_{S_l}$ \end{tabular} 
 %Independent noise ($\mN_S = \bigotimes_{l=1}^n \mN_{S_l}$) 
 & \textcolor{gray}{$\Omega\left(\frac{1}{n}\right)$} & $\Omega\left(\frac{1}{\sqrt{n}}\right)$ & $\Omega\left(\frac{1}{\sqrt{n}}\right)$ & $\Omega\left(\frac{1}{\sqrt{n}}\right)$ \\ 
 \hline
\end{tabular}
\end{center}
\centering
\caption{\label{table} Scalings of our lower bounds on $\epsilon$ and $\delta$ for different noise models (see details in Secs.~III and IV of Ref.~\cite{SM}). Random local noise means noise that acts locally on randomly selected different subsystems, while independent noise means noise that acts independently on all subsystems (which is a mixture of global noises). The lower bounds are taken from both \thmref{thm:global-1} and \thmref{thm:global-2}. Here when we show the lower bound on $\epsilon$ (or $\delta$), we assume $\delta$ (or $\epsilon$) is sufficiently small, i.e., so small that the lower bound on $\epsilon$ (or $\delta$) has the worst scaling. There is a gap between the lower bound on $\delta$ from \thmref{thm:global-1} and \thmref{thm:global-2} for random local noise and a gap between the lower bounds on $\epsilon$ from \thmref{thm:global-1} and \thmref{thm:global-2} for independent noise. A potential way to close up the latter one was explored in Sec.~III B2 of Ref.~\cite{SM}, where the charge fluctuation approach is used to derive a new trade-off relation using quantum metrology.}
\end{table}

\end{document}